\pdfoutput=1

\documentclass[apj]{emulateapj}
\usepackage{amsmath}
\usepackage{graphicx}
\usepackage{xcolor}
\usepackage{verbatim}
\usepackage{mathtools}
\usepackage[caption=false]{subfig}

\shortauthors{Herman et al.}

\begin{document}

\title{Spin-Orbit Misalignment and Precession in the \emph{Kepler}-13A\MakeLowercase{b} Planetary System}

\author{Miranda K. Herman$^{1}$, Ernst J. W. de Mooij$^{2,3}$, Chelsea X. Huang$^{4}$, and Ray Jayawardhana$^{5}$}
\affil{$^{1}$Astronomy \& Astrophysics, University of Toronto, 50 St. George St., Toronto, ON M5S 3H4, Canada;  herman@astro.utoronto.ca}
\affil{$^{2}$School of Physcial Sciences and Centre for Astrophysics \& Relativity, Dublin City University, Glasnevin, Dublin 9}
\affil{$^{3}$Astrophysics Research Centre, Queen's University Belfast, Belfast, United Kingdom BT7 1NN}
\affil{$^{4}$MIT Kavli Institute, 77 Massachusetts Ave, 37-241, Cambridge, MA, USA 02139}
\affil{$^{5}$Physics \& Astronomy, York University, Toronto, Ontario L3T 3R1, Canada}


\begin{abstract}

Gravity darkening induced by rapid stellar rotation provides us with a unique opportunity to characterize the spin-orbit misalignment of a planetary system through analysis of its photometric transit. We use the gravity-darkened transit modeling code \texttt{simuTrans} to reproduce the transit light curve of \emph{Kepler}-13Ab by separately analyzing phase-folded transits for 12 short-cadence \emph{Kepler} quarters. We verify the temporal change in impact parameter indicative of spin-orbit precession identified by \citet{Szabo12} and \citet{Masuda15}, reporting a rate of change $db/dt = (-4.1 \pm 0.2) \times 10^{-5}$ day$^{-1}$. We further investigate the effect of light dilution on the fitted impact parameter and find that less than 1\% of additional light is sufficient to explain the seasonal variation seen in the \emph{Kepler} quarter data. We then extend our precession analysis to the phase curve data from which we report a rate of change $db/dt = (-3.2 \pm 1.3) \times 10^{-5}$ day$^{-1}$. This value is consistent with that of the transit data at a lower significance and provides the first evidence of spin-orbit precession based solely on the temporal variation of the secondary eclipse.

\end{abstract}


\section{Introduction}\label{sec:Intro} 

For about a third of measured exoplanets, the orbital axis is tilted relative to the stellar spin axis by more than two standard deviations \citep{Albrecht12}, a situation referred to as spin-orbit misalignment. In nearly all of these cases the exoplanets are hot Jupiters, which challenges our understanding of how planets form, migrate, and evolve. Studying such misaligned systems is therefore imperative if we wish to understand the origin and dynamical history of systems that differ remarkably from our own.

The Rossiter-McLaughlin (RM) effect \citep{Holt1893, Schlesinger1910} has traditionally been the primary method used to probe spin-orbit misalignment in exoplanetary systems. This tool relies on the distortion of stellar line profiles that occurs when a transiting body eclipses part of the rotating star, and is commonly used in the study of eclipsing binary stars \citep[e.g.][]{Rossiter24, McLaughlin24}. However, the RM approach has two significant drawbacks. First, it can only constrain the sky-projected spin-orbit misalignment $\lambda$. It cannot provide information about the stellar inclination $i_{*}$ or orbital inclination $i_{orb}$, so the true spin-orbit misalignment in 3D-space $\psi$ cannot be determined without other means \citep[e.g.][]{Winn07}. Second, the RM effect is far less precise for hot, rapidly rotating stars. This is because hot stars generally have few spectral lines, and these lines will be significantly broadened by rapid stellar rotation \citep{Johnson14}.

Yet rapid stellar rotation lends itself to spin-orbit determination through transit photometry alone, as shown by \citet{Barnes09}. Fast rotation results in stellar oblateness which induces a gradient in the surface gravity $g$ of the star. \citet{VonZeipel24} describes the effect this has on the effective temperature (and hence emitted flux) at each point on the stellar surface,
\begin{equation} \label{eq:gd_temp}
    \frac{T}{T_{pole}} = \bigg(\frac{g}{g_{pole}}\bigg)^{\beta}
\end{equation}
where the exponent $\beta$ characterizes the strength of this gravity darkening phenomenon and is theoretically 0.25 for a star with a radiative envelope. If the planet's orbit is misaligned with respect to the star's axis of rotation, the planet's transit chord will pass over a nonuniform stellar surface, resulting in an asymmetric transit light curve. This gravity-darkening technique provides an advantage over the RM effect since it can measure both $\lambda$ and $i_{*}$ simultaneously based on the full shape of the transit anomaly.

\emph{Kepler}-13Ab was the first planetary system in which spin-orbit misalignment was identified using transit photometry alone \citep{Szabo11, Barnes11}. A number of groups have since investigated the transit asymmetries it presents \citep[e.g.][]{Szabo14, Masuda15, Howarth17}. Both \citet{Barnes11} and \citet{Masuda15} implement gravity darkened transit models \citep[based on][]{Barnes09} to characterize the spin orbit misalignment, while \citet{Howarth17} employ a model described by \citet{Espinosa11}, approximating the rotational distortion of the host star's surface with a Roche equipotential. There are a number of important factors to consider when comparing these studies. As highlighted by \citet{Howarth17}, the effective temperature of \emph{Kepler}-13A has reported values between 7650-9107 K \citep{Brown11, Szabo11, Huber14, Shporer14}, the lower end of which places the star near the boundary between radiative and convective envelope regimes. This makes it unclear how strongly gravity darkening affects stars like \emph{Kepler}-13A; for convective envelopes $\beta$ is theoretically just 0.08 \citep{Lucy1967}. The intensity of gravity darkening is also influenced by stellar mass: low mass stars show more exaggerated gravity darkening or higher $\beta$ values \citep{Barnes11}. However, the reported mass of \emph{Kepler}-13A is also poorly constrained, with values ranging from 1.72-2.47 $M_{\odot}$ \citep{Borucki11, Szabo11, Shporer14, Huber14}. This provides little clarity as to which stellar envelope regime and $\beta$ value most accurately describe the star. The best-fit parameter values from previous gravity darkened transit models -- and those presented herein -- should therefore be interpreted with these factors in mind. 

\emph{Kepler}-13Ab has also been the subject of some disagreement between literature values of the sky-projected spin-orbit misalignment $\lambda$. Doppler tomography performed by \citet{Johnson14} was inconsistent with the results of \citet{Barnes11}, but \citet{Masuda15} reconciled the two by adjusting the quadratic limb darkening parameters and fixing $\lambda$ to the later spectroscopic measurement. \citet{Howarth17} find a similar solution with their model.

Moreover, not only does the transit show asymmetries characteristic of a spin-orbit misaligned system, but it is also known to vary in duration over the length of the \emph{Kepler} observations. This is indicative of spin-orbit precession caused by the rapidly rotating star's quadrupole moment \citep{Szabo12, Szabo14, Masuda15}. The precession can very clearly be seen in the temporal variation of the impact parameter between transits.

To date no evidence of this variation has been seen in the phase curve and secondary eclipse of the planet, though given the amplitude of the variability and the sublime quality of the \emph{Kepler} data, this possibility is worth exploring. In a system not unlike \emph{Kepler}-13Ab, \citet{Armstrong16} recently presented the first detection of variations in the peak offset of a planet's phase curve. HAT-P-7b, a hot Jupiter, showed signs of atmospheric variability indicative of changing cloud coverage -- something only seen previously in the atmospheres of brown dwarfs \citep[e.g.][]{Radigan12, Wilson14, Rajan15}. 

In this paper, we perform an entirely independent analysis of the transit data of \emph{Kepler}-13Ab by modeling the binned and phase-folded transit light curves for 12 short-cadence quarters, determining both the spin-orbit misalignment and precession rate. We also analyze the secondary eclipse in an effort to further confirm the change in impact parameter seen in the transit data.


\begin{deluxetable}{cccc}
\tablecolumns{4}
\tablewidth{0pt}
\tablecaption{\emph{Kepler} Data Quarters Used in Analysis}
\tablehead{
\colhead{Quarter}    &  \colhead{Cadence}  &  \colhead{Transits Used}  &  \colhead{Phase Curves Used}
}
\startdata
0                   & l     & \nodata  & 6                \\
1                   & l     & \nodata  & 19               \\
2                   & s     & 46       & 50            \\
3                   & s     & 44       & 47              \\
4                   & l     & \nodata  & 49               \\
5                   & l     & \nodata  & 52               \\
6                   & l     & \nodata  & 49               \\
7                   & s     & 47       & 50              \\
8                   & s     & 34       & 37                     \\
9                   & s     & 51       & 54                     \\
10                  & s     & 49       & 52                     \\
11                  & s     & 49       & 52                     \\
12                  & s     & 41       & 41                     \\
13                  & s     & \nodata  & 50                     \\
14                  & s     & 47       & 51                     \\
15                  & s     & 47       & 50                     \\
16                  & s     & 39       & 41                     \\
17                  & s     & 13       & 15                    
\enddata
\tablenotetext{}{{\bf Note}: The `s' and `l' indicate short- and long-cadence, respectively. Only short-cadence data were used in the transit analysis.}
\label{tab:quarters}
\end{deluxetable}

\section{Data Reduction} \label{sec:Data}
We use the \emph{Kepler} simple aperture photometry (SAP) data outlined in Table \ref{tab:quarters} for our analysis of the light curve. For the transit we use the short-cadence (60 second exposure) data, while for the phase curve we use both long- and short-cadence data, binning the latter into 30-minute bins to match the cadence of the former. We do not include the long-cadence data in our transit analysis since it tends to smear out important features and would consequently reduce the precision of our fitted transit model. We discard any bad data points flagged by \emph{Kepler} in the SAP files and normalize the data to the out-of-transit median. We exclude quarter 13 from our analysis because the standard deviation of its raw out-of-transit data was much higher than other quarters.

\subsection{Companion Flux Contribution} \label{sec:Companion}
Because \emph{Kepler}-13A is part of a binary system \citep{Aitken1904} that is unresolved in the \emph{Kepler} data, we must account for the additional flux its companion contributes to the light curve. The reported amount by which the light curve is diluted varies among studies, from 38--48\%  \citep{Szabo11,Adams13,Shporer14}. We use the value from \citet{Shporer14} in our analysis, removing 48\% of the light in each SAP file.

\subsection{Removal of Phase Variations and Systematics} \label{sec:Systematics}
We then remove the following out-of-transit variations from the data in each SAP file: 
\begin{itemize}
    \item Variations due to Doppler boosting and ellipsoidal signals, $F_{m}$, both of which are effects due to the mass of the planet. Doppler boosting is a combined result of brightness increasing along the radial velocity vector and a red/blueshift of the stellar spectrum that shifts parts of the spectrum in and out of the bandpass, influencing the observed brightness. Ellipsoidal variations are due to the tidal bulge raised on the stellar surface as the planet orbits, changing the observed surface area of the star.
    \item The brightness contribution of the planet, $F_{p}$, which varies as the illuminated side of the planet moves in and out of view.
    \item The third harmonic signal in the planet's period, $F_{3}$, identified in \citet{Esteves13} and \citet[][from here on E15]{Esteves15} but not yet well understood in terms of the underlying cause. They suggest that it could be due to the spin-orbit misalignment presenting as a perturbation of the ellipsoidal variations.
\end{itemize}

Notably, all of these variations depend on orbital phase $\phi$, and we refer the reader to E15 for the detailed equations describing them. We remove the variations from the normalized out-of-transit flux such that
\begin{equation}
    F'(\phi) = \frac{F(\phi) - F_{p}(\phi)}{1 + F_{m}(\phi) + F_{3}(\phi - \theta_{3})}
\end{equation}
where $\phi$ runs from 0 to 1 with mid-transit at $\phi=0$, and $\theta_{3}$ is the cosine third harmonic phase offset. For $\theta_{3}$ and other constants used to calculate the phase variations, we use the values from Table 8, Model 1 in E15.

We then calculate the out-of-transit median and remove any outliers above 3$\sigma$. To remove instrumental signals we use a least squares fit of a third order polynomial to $\pm0.2$ days around the transit, and divide the variation-subtracted SAP flux in this phase range by the resulting fit. We elected to use this method rather than fit and divide out the \emph{Kepler} cotrending basis vectors (CBVs) in order to avoid complications introduced by the chosen number of CBVs fitted and their relative contributions to the systematic signal.

\begin{figure}
	\centering
   \includegraphics[width=0.49\textwidth]{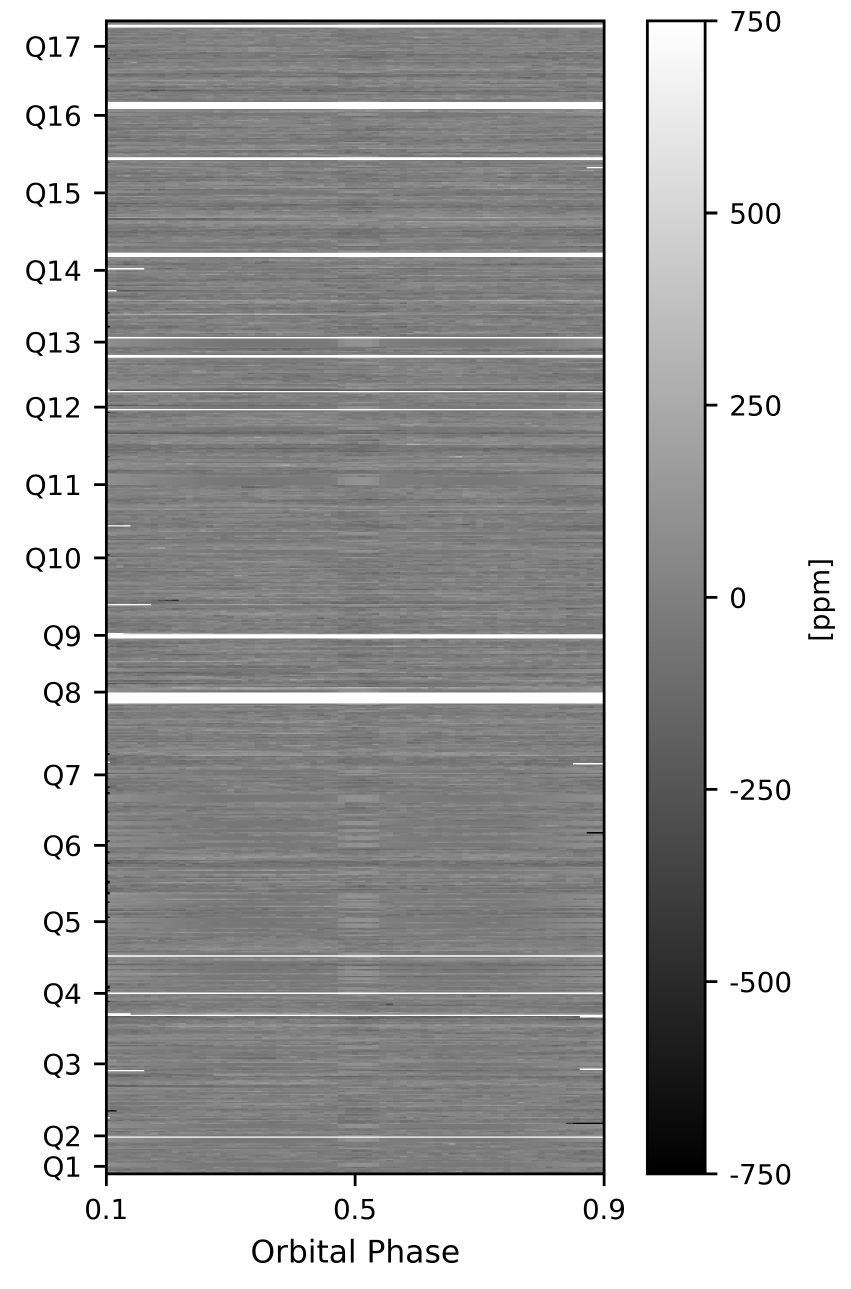}
    \caption{A waterfall plot displaying the difference between each phase curve and the mean overall phase curve. We use the reduced long- and short-cadence data from E15 and bin all data into 30-minute bins. Ingress and egress of the secondary eclipse fit into single bins, meaning they are strongly affected by discrepancies between phase curves, but there could be variability at these locations. White horizontal lines correspond to data gaps. 
    } 
    \label{fig:phase_waterfall}
\end{figure}

\subsection{Transit Phase-Folding and Binning} \label{sec:Binning}
After removing companion dilution, variations, and systematics from the SAP flux in each file, we cut all light curves to $\pm0.15$ days around each transit so we may focus on the asymmetries and variations in the transit alone. We then phase-fold the transit light curves for each \emph{Kepler} quarter and bin the data points into one-minute bins. 

While some previous studies have focused on analyzing a single, binned, phase-folded light curve that incorporates data from all available \emph{Kepler} transit events at their respective times of publication \citep[i.e.][]{Barnes11,Szabo11,Esteves13,Esteves15,Howarth17}, by separately analyzing the binned and phase-folded light curves for each quarter we are able to determine any changes in the transit parameters over time. Phase-folding over entire quarters, rather than analyzing individual transit events like \citet{Szabo12} or \citet{Masuda15}, has the benefit of increasing the signal-to-noise ratio of the transits which allows for more precise model fitting while preserving most of the gradual time-dependent variation.

\begin{figure}
	\centering
   \includegraphics[width=0.48\textwidth]{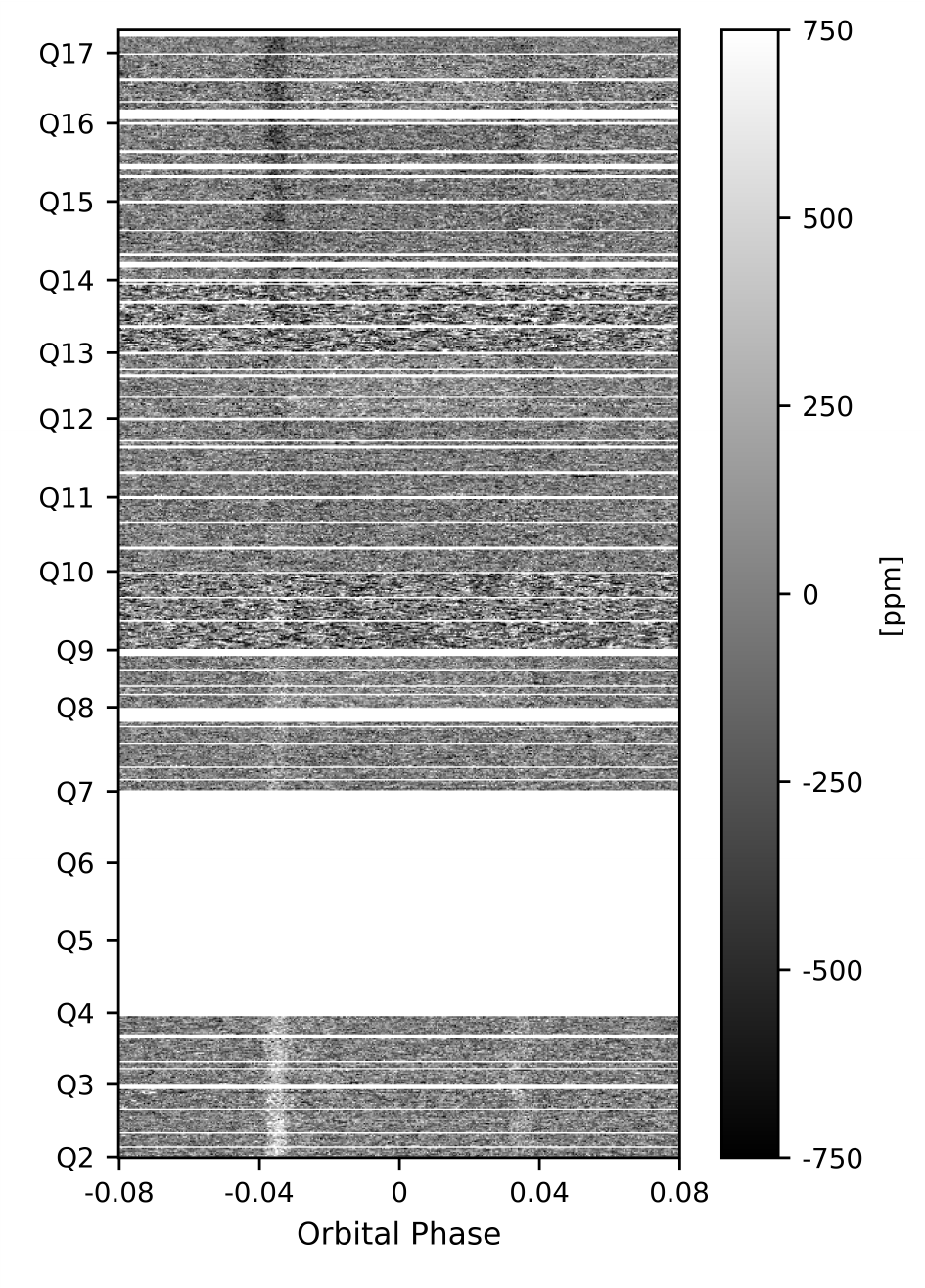}
    \caption{A waterfall plot displaying the difference between each transit and the mean overall transit. The white horizontal lines correspond to gaps in the short-cadence observations. The gradient seen in the two vertical lines near ingress and egress indicates that there is a change in the transit shape: The transit is becoming wider over time.
    }  
    \label{fig:waterfall}
\end{figure}


\section{Qualitative Analysis} \label{sec:qual}

Before performing a quantitative analysis of the phase curve and transit data we consider whether variations are evident from a purely qualitative perspective. Figure \ref{fig:phase_waterfall} shows the difference between each phase curve and the mean overall phase curve. Unlike the transit data, we use the reduced phase curve data from E15 which includes both short and long-cadence data binned into 30-minute bins. The secondary eclipse occurs at $\phi = 0.5$. It is difficult to discern any noticeable gradient in the phase curves by eye, however the boundary of the secondary eclipse suggests the possibility for variability; the entirety of its ingress and egress fit into single bins, so these points are more strongly affected by discrepancies between phase curves. We therefore elect to investigate the secondary eclipse (though with a less rigorous model than that used for the transit data) and we outline this analysis in \S \ref{sec: phase curve}.

\begin{figure}
	\centering
   \includegraphics[width=0.48\textwidth]{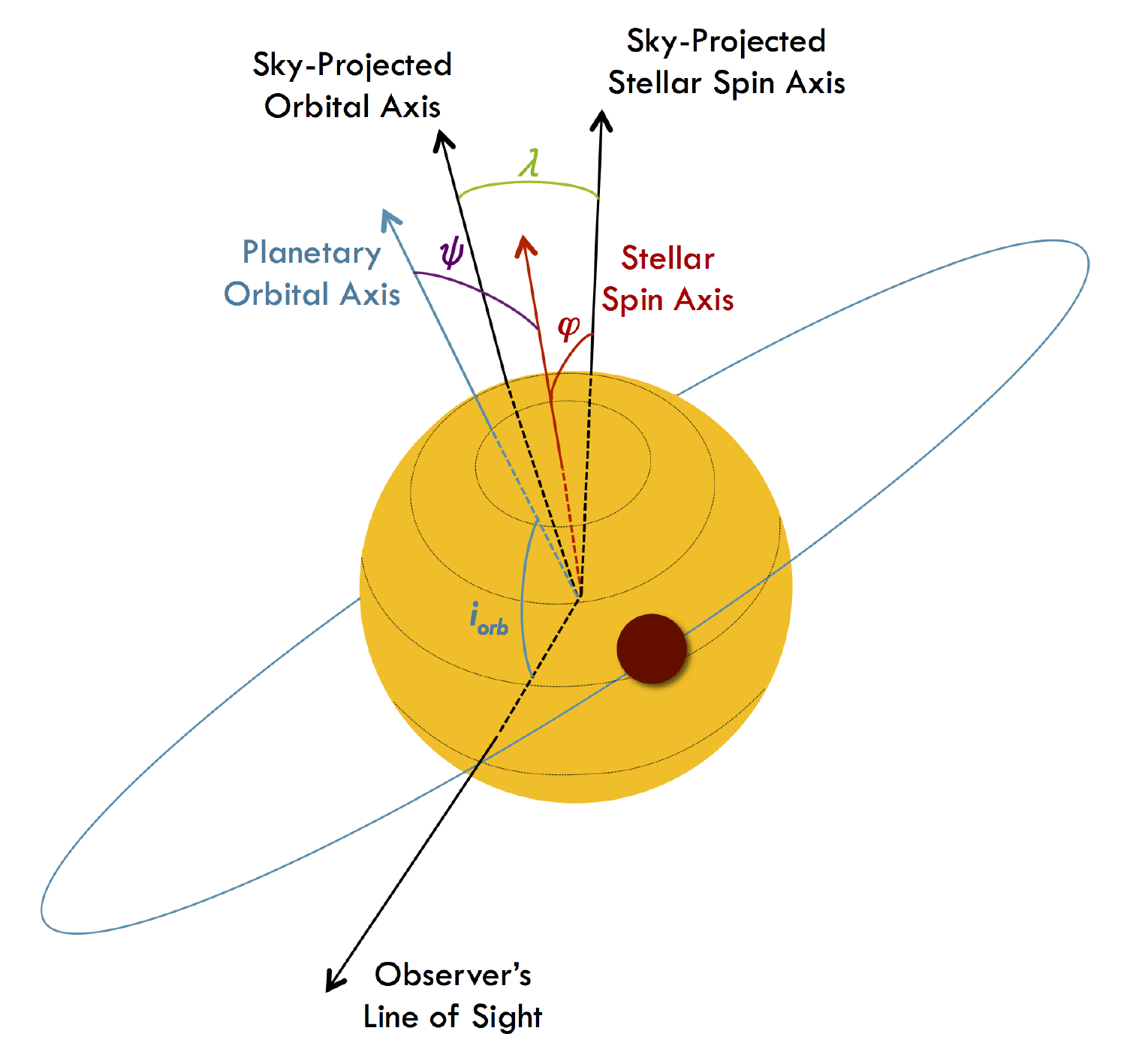}
    \caption{Illustration of the geometry of a spin-orbit misaligned system. $\varphi$ is the angle between the sky plane and the stellar spin axis, while $i_{orb}$ is the angle between the observer's line of sight and the planet's orbital axis. $\psi$ is the true spin-orbit misalignment angle in 3D space and $\lambda$ is its 2D projection onto the sky plane.
    }  
    \label{fig:orbit_params}
\end{figure}

Figure \ref{fig:waterfall} shows the difference between each corrected short-cadence transit and the mean overall transit. Here we find qualitative confirmation of the change in transit shape caused by decreasing impact parameter, as reported by \citet{Szabo12} and \citet{Masuda15}. The vertical gradient seen near ingress and egress indicates that the transit duration is increasing with time; an effect of decreasing impact parameter (and thus orbital inclination). The gradient at egress is noticeably less pronounced than its ingress counterpart. This could be due to the slight variation in the time of mid transit $T_{0}$ we identify in \S \ref{sec:spin-orbit precession}. In the following section we describe our transit model used to quantify the variation we see in the transit data.


\section{Transit Model} \label{sec:model}

We use the modeling code \texttt{simuTrans}\footnote{https://github.com/chelseah/simuTrans}, a numerical integrator created to model transit light curves affected by gravity darkened stars. It is based on the methods of \citet{Barnes09}. The star is defined on a grid with the brightness at each point modeled as blackbody emission for a temperature described by equation \ref{eq:gd_temp}. After setting a series of fixed parameters, \texttt{simuTrans} employs \texttt{emcee}, a Markov Chain Monte Carlo (MCMC) ensemble sampler \citep{Foreman2013} to explore the posterior probability distribution for the remaining fitted parameters and determine their best fit values.

\begin{deluxetable}{cc}
\tablecolumns{2}
\tablewidth{0pt}
\tablecaption{Transit Model Parameters}
\tablehead{
\colhead{Parameter} & \colhead{Value}
}
\startdata
P (days)\tablenotemark{a} & 1.763588  \\
M$_{*}$ (M$_{\odot}$)\tablenotemark{b} & 1.72  \\
R$_{*}$ (R$_{\odot}$)\tablenotemark{b} & 1.74   \\
P$_{rot}$ (hr)\tablenotemark{c} & 22.5   \\
$\beta$ & 0.25  \\
f$_{*}$ & 0.02 $\pm$ 0.01   \\
q$_{1}$ & 0.261 $\pm$ 0.017 \\
q$_{2}$ & 0.350 $\pm$ 0.040  \\
R$_{*}/a$ & 0.2205 $\pm$ 0.0013  \\
R$_{p}$/R$_{*}$ & 0.0905 $\pm$ 0.0004   \\
T$_{o}$ (days)\tablenotemark{d} & 120.56578 $\pm$ 0.000011  \\
b & 0.1960 $\pm$ 0.0013 \\
$\varphi$ ($^{o}$) & 13.0 $\pm$ 0.8 \\
$\lambda$ ($^{o}$)  & -27.9 $\pm$ 1.1 \\
\enddata
\tablenotetext{a}{From E15}
\tablenotetext{b}{From \citet{Shporer14}}
\tablenotetext{c}{From \citet{Barnes11}}
\tablenotetext{d}{{$T_{o}$ is given in BJD -- 2454833 days}}
\label{tab:Params}
\end{deluxetable}

The fixed parameters are:
\begin{enumerate}
    \itemsep0em
    \item Orbital period, $P$
    \item Stellar mass, $M_{*}$
    \item Stellar radius, $R_{*}$
    \item Stellar rotation period, $P_{rot}$
    \item Gravity darkening exponent, $\beta$
    \item Stellar oblateness, $f_{*} = 1 - \frac{R_{pole}}{R_{eq}}$
    \item Limb darkening coefficient, $q_{1} = (u_{1} + u_{2})^{2}$
    \item Limb darkening coefficient, $q_{2} = \frac{u_{1}}{2(u_{1} + u_{2})}$
    \item Stellar radius scaled by semi-major axis, $R_{*}/a$
    \item Planetary radius scaled by stellar radius, $R_{p}/R_{*}$
\end{enumerate}
where $q_{1}$ and $q_{2}$ are the re-parametrized quadratic limb darkening coefficients outlined by \citet{Kipping2013}. We also assume a circular orbit since the eccentricity has been shown to be very small: on the order of $10^{-4}$ \citep{Benomar14,Shporer14,Esteves15}.

The fitted parameters are:
\begin{enumerate}
    \itemsep0em
    \item Time of mid transit, $T_{o}$
    \item Impact parameter, $b$
    \item Angle between sky plane and stellar spin axis, $\varphi$
    \item Sky-projected angle between stellar spin axis and planetary orbital axis, $\lambda$
\end{enumerate}

\begin{figure*}
    \centering
    \includegraphics[width=\textwidth]{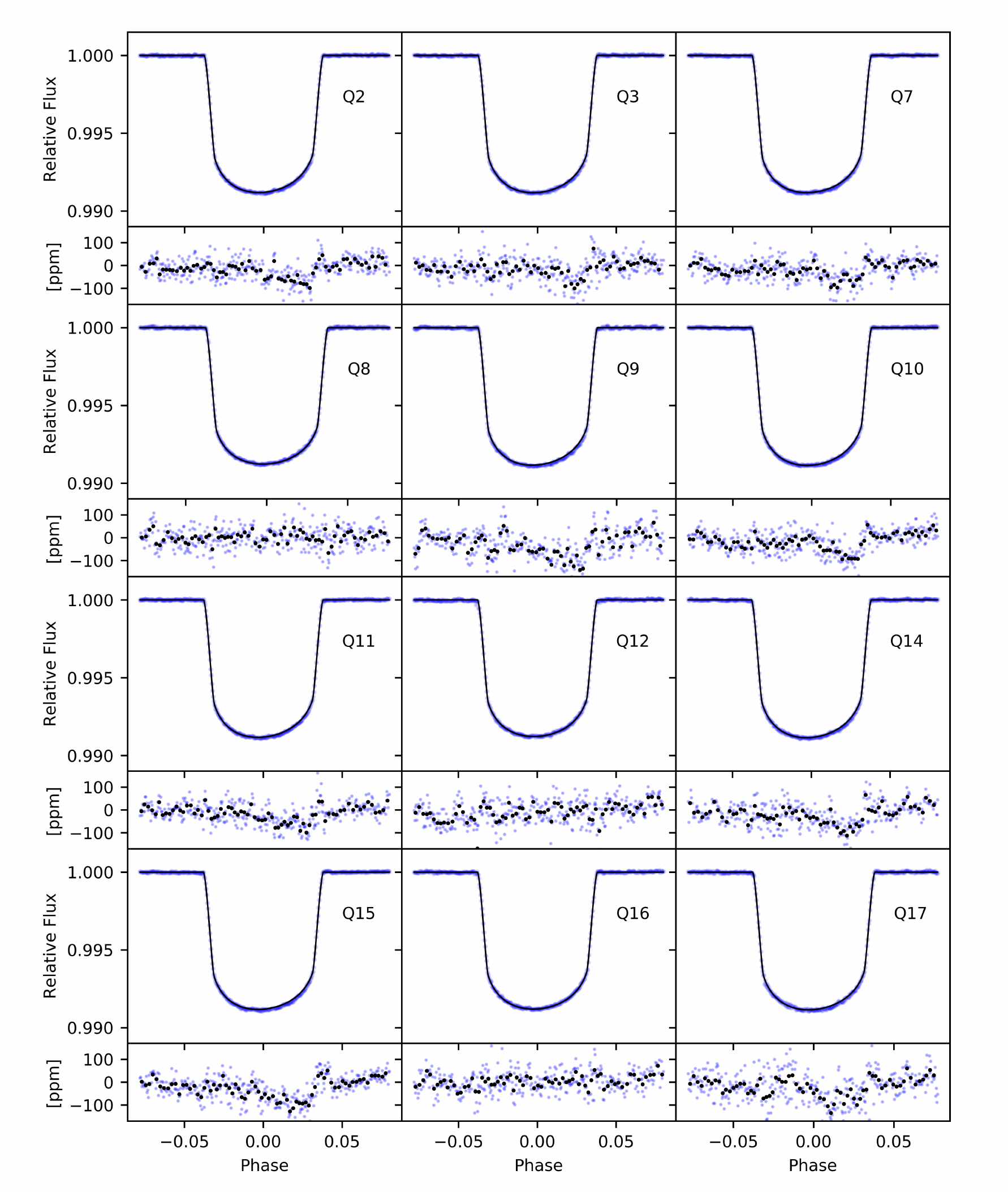}
    \caption{The transits, models, and residuals for all fitted quarters. For each, the upper panel shows the model (solid black line) fitted to the one-minute binned and phase-folded transit (blue circles). The bottom panel shows the residuals (blue) characterizing the difference between the data and the model at each point during the transit. The black over-plotted points have been binned with 5 data points per bin.}
    \label{fig:all_models}
\end{figure*}

While the values for fixed parameters 1 - 5 are taken directly from previous studies (see Table \ref{tab:Params}), the values for  fixed parameters 6 - 10 were determined through an iterative MCMC fit. All of the short-cadence data for \emph{Kepler}-13Ab were divided into four groups based on their year of observation. These were then phase-folded into four transits following the same method described in \S \ref{sec:Data}, and then fitted with \texttt{simuTrans}. Including the four previously listed fitted parameters, we fit for nine parameters total. We expected parameters 6 - 10 to remain constant, since the limb-darkening coefficients describe an intrinsic property of the star and $R_{*}/a$ and $R_{p}/R_{*}$ are intrinsic to the planetary system. The parameter $f_{*}$ is related to the star's rotational velocity, which should not change over the timescales considered. Therefore, after discerning that these parameters remained constant throughout, we calculated their average values and set these as fixed parameters when fitting the quarter-folded light curves. The values reported in Table \ref{tab:Params} for the four fitted parameters are the mean values calculated from all of the quarter-folded transit fits.

Figure \ref{fig:orbit_params} shows the definitions of the various angles used to describe the configuration of the system. The angle of the stellar spin axis $\varphi$ is measured from the sky plane down along the observer's line of sight (LOS), and is defined to be in the range [$-90^{o}$, $90^{o}$]\footnote{Note that our definition of $\varphi$ differs by $90^{o}$ from $i_{*}$ in \citet{Masuda15}, which is defined from the observer's LOS upward.}.
The angle from the sky-projected planetary orbital axis to the sky-projected stellar spin axis, $\lambda$, is measured on the sky plane perpendicular to the LOS. It is defined to be in the range [$-180^{o}$, $180^{o}$]. To determine the true spin-orbit misalignment $\psi$ in 3D space (see Figure \ref{fig:orbit_params}), the planet's orbital inclination must also be taken into account: 
\begin{equation} \label{eq: cosi}
    \cos i_{orb} = \Big(\frac{R_{*}}{a}\Big) b.
\end{equation}
The true spin-orbit misalignment can then be found by modifying equation (1) of \citet{Benomar14} to match our parameter definitions such that
\begin{equation} \label{eq: psi}
    \cos \psi = \sin\varphi \cos i_{orb} + \cos\varphi \sin i_{orb} \cos \lambda.
\end{equation}


\section{Results and Discussion} \label{sec:results}

In this section, we examine the results of the models fitted to the phase-folded and binned transits for the 12 quarters considered, as well as the implications on the future evolution of \emph{Kepler}-13Ab. The fitted models are shown in Figure \ref{fig:all_models}. There is a slight asymmetry in the residuals during egress in some of the models which could be due to the fixed value of the gravity darkening parameter $\beta$. As discussed in \S \ref{sec:Intro}, it is unclear whether the host star is best described by a convective or radiative stellar envelope, and exploring the resultant influence on $\beta$ lies outside the scope of this paper. We leave this work for future consideration and instead adopt the value $\beta = 0.25$ which is consistent with previous studies of this system \citep[e.g.][]{Barnes11,Masuda15}.

\begin{figure}
	\centering
   \includegraphics[width=0.48\textwidth]{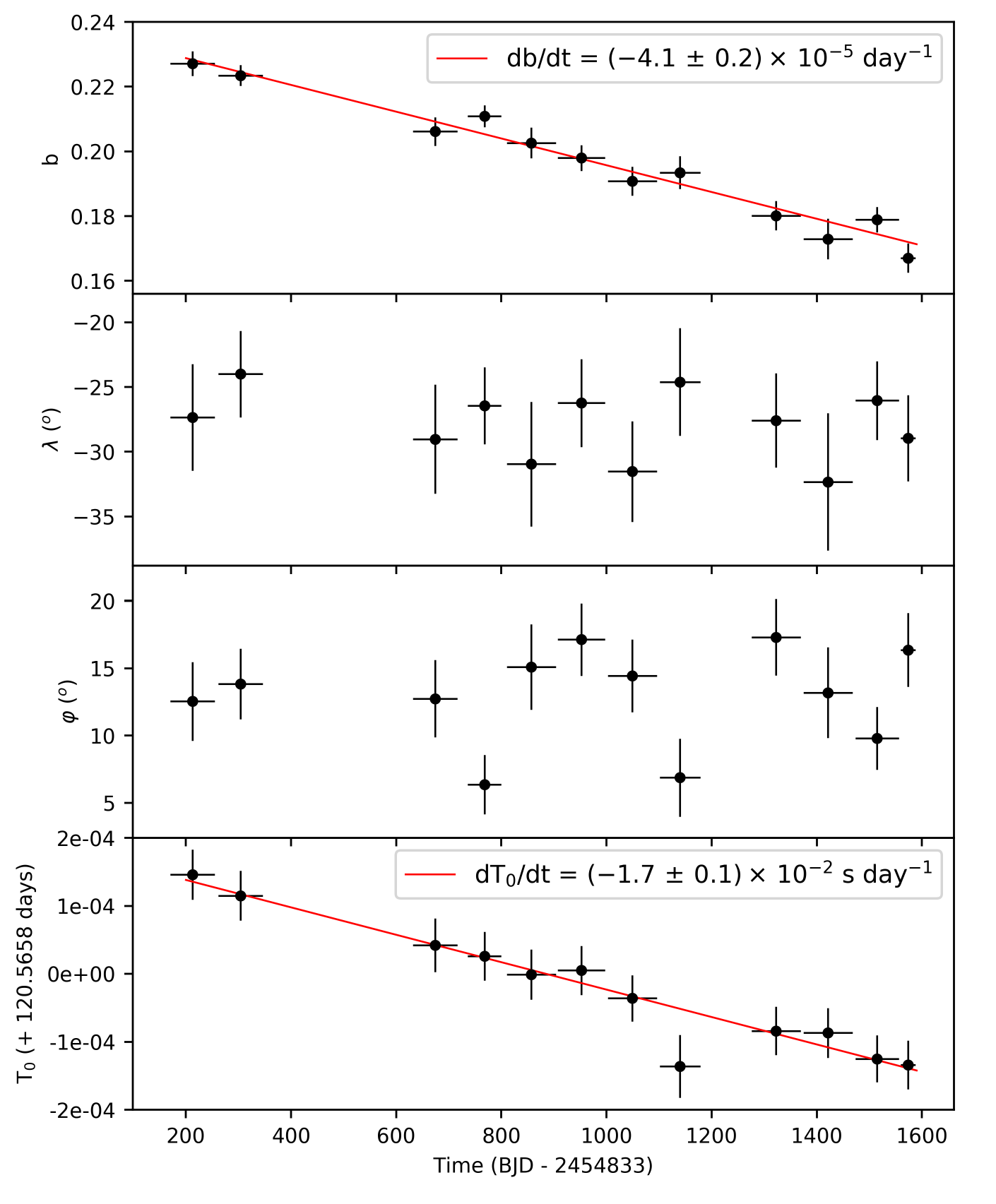}
    \caption{\emph{Top Panel:} Evolution of the fitted impact parameter over time. Each point corresponds to a separate short-cadence \emph{Kepler} quarter. The red line is a least squares fit from which we report the rate of change of b. The horizontal error bars show the time-span of each quarter. 
    \emph{Second Panel:} Fitted values of the sky-projected spin-orbit misalignment $\lambda$ for all quarters.
    \emph{Third Panel:} Fitted values of the stellar inclination $\varphi$ for all quarters.
    \emph{Bottom Panel:} Evolution of the fitted time of mid transit over time. The red line shows a least squares fit from which we find a very small rate of change. 
    }  
    \label{fig:b_evolution}
\end{figure}

\subsection{Spin-Orbit Precession} \label{sec:spin-orbit precession}

The evolution of the fitted parameters $b$, $\lambda$, $\varphi$, and $T_{0}$ are plotted in Figure \ref{fig:b_evolution}. The top panel shows a clear decrease in the impact parameter over the time span of the short-cadence data quarters, which indicates that the inclination of \emph{Kepler}-13Ab's orbit is undergoing temporal variation.  We fit a linear model to this time series using a least squares method and determine a rate of change of $db/dt = (-4.1 \pm 0.2) \times 10^{-5}$ day$^{-1}$. This is consistent with previously reported values; \citet{Szabo12} find $db/dt = (-4.4 \pm 1.2) \times 10^{-5}$ day$^{-1}$ by individually fitting each transit in \emph{Kepler} quarters 2 and 3, while \citet{Masuda15} fit individual transits in all short-cadence quarters (including quarter 13) and find $d|\cos(i_{orb})|/dt = (-7.0 \pm 0.4) \times 10^{-6}$ day$^{-1}$. Equation \ref{eq: cosi} gives the relationship between $\cos i_{orb}$ and $b$.

The consistency between these results is reassuring considering the differences in our methods. We use phase-folded transit light curves which considerably improve the signal to noise ratio of the data compared to individual transits, but do not wash out the clear trend in impact parameter over the timescale considered. We also utilize more than 6 times as much data as \citet{Szabo12} by including \emph{Kepler} quarters 7-12 and 14-17 in our analysis.

The temporal variation of the impact parameter has important implications, as first established by \citet{Szabo12} and confirmed by both \citet{Masuda15} and this work. Namely the variations are indicative of spin-orbit precession caused by the host star's quadrupole moment. Based on the rate of change in impact parameter we expect \emph{Kepler}-13Ab to become a non-transiting exoplanet in 75 - 85 years.

We also find a small rate of change in the time of mid transit $T_{0}$ of $(-1.7 \pm 0.1) \times 10^{-2}$ s $\cdot$ day$^{-1}$, as shown in the bottom panel of Figure \ref{fig:b_evolution}. This is most likely not caused by spin-orbit precession and instead likely results from the value at which we fix the orbital period (see Table \ref{tab:Params}). The decrease in $T_{0}$ suggests that this period is too large by $3.6 \times 10^{-7}$ days, or about 0.03 seconds. Beyond this small correction we do not find any evidence for a temporal change in the period at the significance level of our measurements.

\subsection{Spin-Orbit Misalignment} \label{sec:spin-orbit misalignment}

Since the angular momenta of the stellar spin and orbital motion have comparable magnitudes in the \emph{Kepler}-13A system (due to the star's rapid rotation and the planet's small semi-major axis), they will both precess about the total angular momentum vector. As a result the angles $i_{orb}$, $\lambda$, and $\varphi$ are all expected to undergo variation. \citet{Masuda15} computes the future evolution of these parameters and shows that neither $\lambda$ nor $\varphi$ should change by more than a few degrees on the relatively short timescale we consider. Indeed, the middle and bottom panels of Figure \ref{fig:b_evolution} exhibit no discernible temporal variation within error.

\begin{figure}
	\centering
   \includegraphics[width=0.474\textwidth]{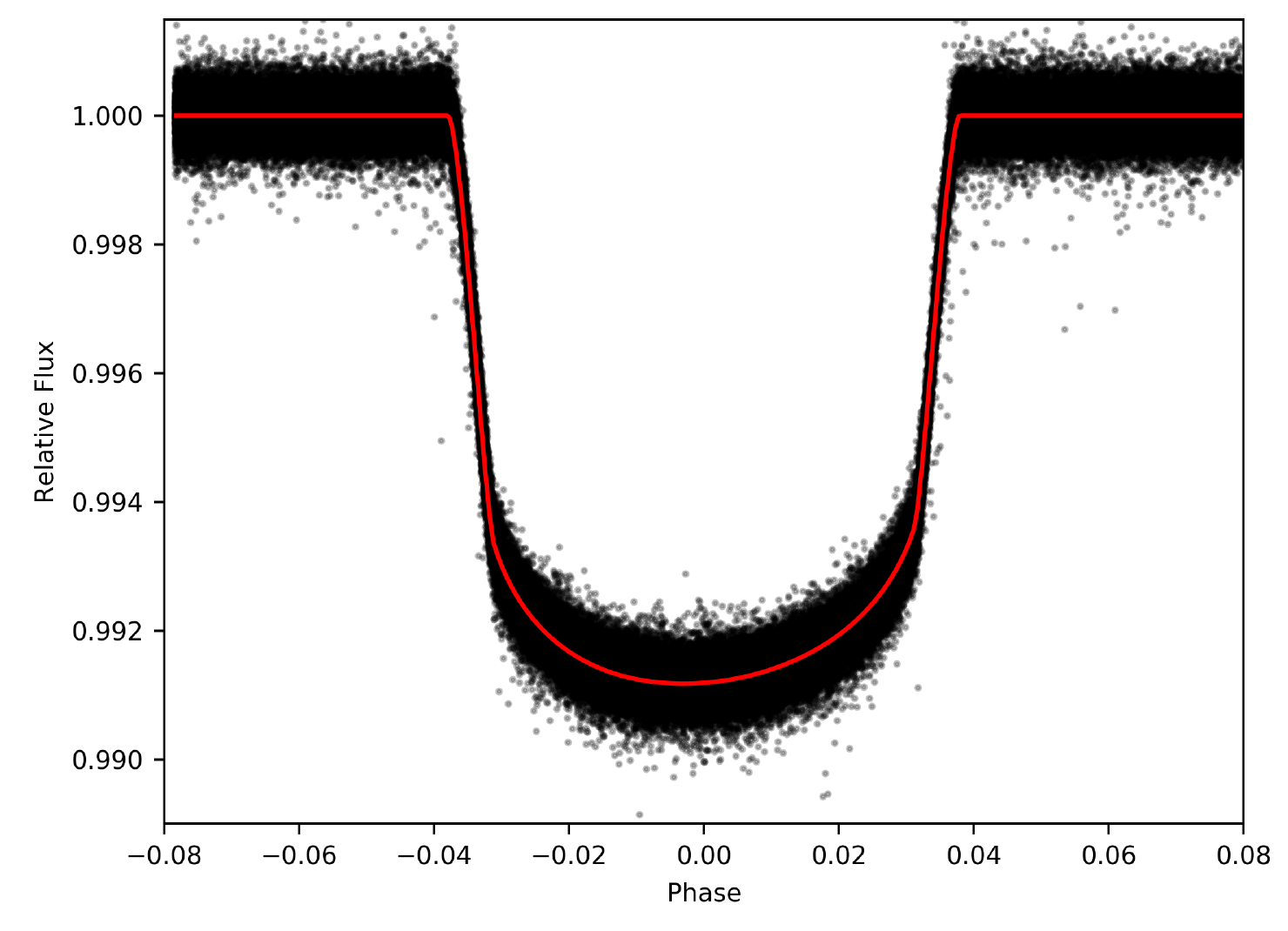}
    \caption{The average model (solid red line) computed using the mean fitted parameters reported in Table \ref{tab:Params}, plotted on top of the phase-folded transit data (black dots).
    }  
    \label{fig:avg_model}
\end{figure}

We therefore use the average values of $b$, $\lambda$, and $\varphi$ to calculate the true spin-orbit misalignment $\psi$ using equations \ref{eq: cosi} and \ref{eq: psi}. We report a misalignment of $29^{o} \pm 1^{o}$ for this system. Our value is lower than that given in \citet{Masuda15}, likely due to the difference in our fitted value of $\lambda$. In Figure \ref{fig:avg_model} we use the average values of our fitted parameters to plot the mean transit model on top of a single phase-folded transit encompassing all transits in the 12 short-cadence quarters investigated. 

\begin{figure}
	\centering
   \includegraphics[width=0.478\textwidth]{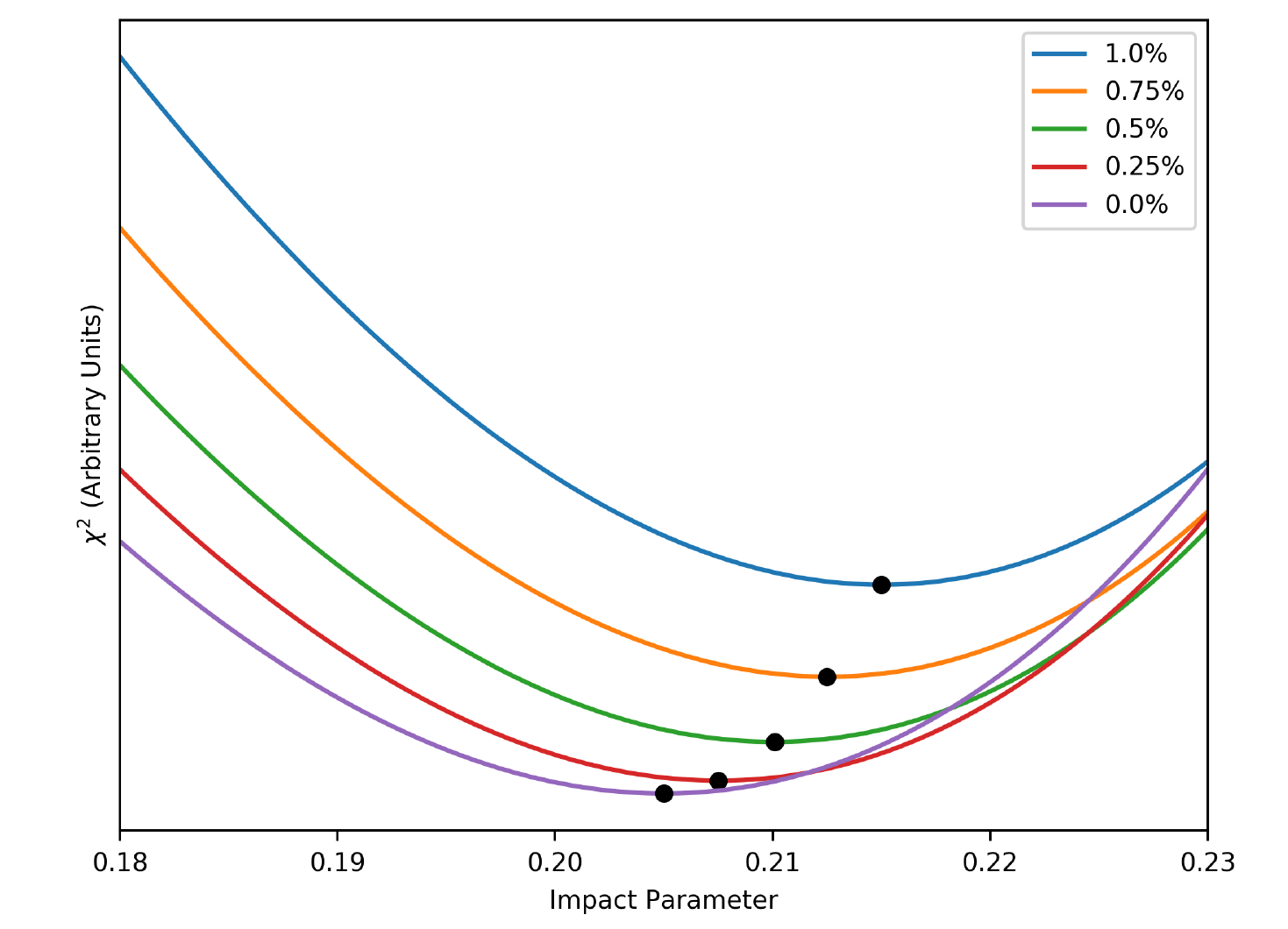}
    \caption{Effect of contaminating light on fitted impact parameter. If the dilution is not accounted for, the best-fit model (based on the minimum $\chi^{2}$) will have a larger impact parameter to compensate for the shallower transit depth. The black dots denote the minimum $\chi^{2}$ values, indicative of the shift in impact parameter.
    }  
    \label{fig:chisq_v_b}
\end{figure}

\subsection{Seasonal Variations} \label{sec:Evolution}

The top panel of Figure \ref{fig:b_evolution} contains a few systematically offset values which are from quarters 8, 12, and 16. These three quarters correspond to the same `season' in each year of the \emph{Kepler} observations meaning the telescope is rotated onto the same side and \emph{Kepler}-13A is located on the same CCD each time. Because of this, the offset has previously been attributed to systematic rather than physical variation, both for \emph{Kepler}-13Ab \citep{Masuda15} and other planetary systems such as HAT-P-7b \citep{VanEylen13}. \citet{VanEylen13} suggest instrumental artifacts or unknown field crowding could be responsible which would translate to an increase in flux that could contaminate the star. Much like the flux contributed by \emph{Kepler}-13A's stellar companion, this has the effect of making the transit shallower, and if unaccounted for will result in a larger impact parameter when fitting the gravity-darkened transit model.

To investigate the effect of flux dilution on the fitted impact parameter, we produce and compare quadratically limb-darkened transit models (without gravity-darkening) for various percentages of contaminating light and a range of impact parameter values. We use the quadratic limb-darkening model from \citet{MandelAgol02}, as implemented with \texttt{Python} in the \emph{occultquad} function\footnote{http://astroutils.astronomy.ohio-state.edu/exofast} from \texttt{EXOFAST} \citep{Eastman12}. We use the same values of $q_{1}$, $q_{2}$, $R_{p}/R_{*}$, and $R_{*}/a$ as those in our gravity-darkened model (see Table \ref{tab:Params}), changing only the impact parameter and percentage of contaminating light. Excluding gravity darkening from this analysis provides a less robust but sufficiently accurate transit model for our investigation.

We first produce a quadratically limb-darkened transit model with zero additional light and a fixed impact parameter value of 0.205, which is approximately the value we would expect for quarter 8 based on the fit shown in Figure \ref{fig:b_evolution}. We then calculate new diluted transit models with additional light between 0 and 1\% and vary the impact parameter in the range [0.18, 0.23] for each. We introduce random noise to the data, with standard deviation equal to the median out-of-transit standard deviation of the transits phase-folded by quarter. We then calculate the $\chi^{2}$ value with respect to the initial model with fixed impact parameter and zero dilution. 

Our results are presented in Figure \ref{fig:chisq_v_b}, with black dots indicating the minimum $\chi^{2}$ for each dilution percentage. From this we find that a percentage of contaminating light less than 1\% is enough to cause the jump in impact parameter seen in quarters 8, 12, and 16, supporting the idea that systematic effects are indeed the cause of such seasonal variations. Determining the precise percentage of contamination in these quarters exceeds the scope of this paper, however, and we treat the offset values as scatter while noting their likely origin.

\begin{deluxetable}{cc}
\tablecolumns{2}
\tablewidth{0pt}
\tablecaption{Phase Curve Results}
\tablehead{
\colhead{Parameter} & \colhead{Value}
}
\startdata
F$_{ecl}$ (ppm) &172.4  $\pm$ 0.7   \\
db/dt & $(-3.2 \pm 1.3) \times 10^{-5}$ day$^{-1}$ \\
dT$_{0,ecl}$/dt & $(-1.7 \pm 0.7) \times 10^{-1}$ s $\cdot$ day$^{-1}$ \\
\enddata

\label{tab:eclipse_params}
\end{deluxetable}

\subsection{Secondary Eclipse Analysis} \label{sec: phase curve}

Despite strong evidence of spin-orbit precession from the \emph{Kepler} transit data, it has not yet been ascertained whether similar variations are perceptible in the phase curve data. We therefore analyze the secondary eclipse as a function of time and describe our findings herein.

Taking into account the appropriate time binning, we model the secondary eclipses of all short- and long-cadence data. We use the same \emph{occultquad} code described in \S \ref{sec:Evolution}, with no limb darkening or gravity darkening. We implement a simple MCMC process to fit for the secondary eclipse depth $F_{ecl}$ for all phase curves, then perform a joint fit assuming constant depth and allowing the impact parameter to vary linearly with time. We also allow for a phase offset to account for the slight deviation in the time of mid eclipse identified in the transit data (see \S \ref{sec:spin-orbit precession}). Our results are presented in Table \ref{tab:eclipse_params}. Though at a much lower significance, the change in impact parameter we measure here is consistent with the value we report from the transit data.

This is the first time confirmation of spin-orbit precession has been identified in the secondary eclipse data. This new result serves as further evidence of the unique capabilities of \emph{Kepler} owing to its long baseline and high photometric precision.


\section{Conclusions} \label{sec:Conclusions}

The prospect of asymmetric transit light curves resulting from spin-orbit misaligned planets orbiting gravity darkened stars was initially discussed by \citet{Barnes09}. \citet{Szabo11} were the first to show that \emph{Kepler}-13Ab displays signs of an asymmetric light curve, and \citet{Barnes11}, \citet{Masuda15}, and \citet{Howarth17} each determined the net spin-orbit misalignment of the system using independent transit models. Transit duration variations due to precession of the planetary orbital axis were identified by \citet{Szabo12} and confirmed by \citet{Masuda15}.

Through analysis of the phase-folded transit light curves for 12 short-cadence data quarters, as well as the individual secondary eclipses, we similarly conclude that the detected change in impact parameter is indicative of a precessing planetary orbit and determine the net spin-orbit misalignment of the system. The calculated rates of change we report based on the transit and phase curve are self-consistent and agree with previously published values. We posit that the systematic offset in quarters 8, 12, and 16 identified in both our light curve analysis and that of \citet{Masuda15} can be explained by the addition of contaminating light from either instrumental artifacts or field crowding, as described by \citet{VanEylen13}. 

Though we are able to use a gravity-darkened transit model to characterize this system, the origin of its misalignment is not yet fully understood, and discerning its complex evolution will require additional short-cadence, high-precision observations like those possible with the upcoming \emph{TESS} mission. Similar observations of other planetary systems like \emph{Kepler}-13Ab will also provide insight into the prevalence of such fascinating dynamical phenomena.

\acknowledgements
We thank Lisa Esteves for her guidance during the early stages of this work, as well as Dan Tamayo for helpful discussions. This work was supported in part by grants from the Natural Sciences and Engineering Research Council (NSERC) of Canada to R.J. E.d.M. was in part funded by the Michael West Fellowship.

\bibliographystyle{apj}
\bibliography{bibliography}

\begin{thebibliography}{}
\expandafter\ifx\csname natexlab\endcsname\relax\def\natexlab#1{#1}\fi

\bibitem[{{Adams} {et~al.}(2013){Adams}, {Ciardi}, {Dupree}, {Gautier},
  {Kulesa}, \& {McCarthy}}]{Adams13}
{Adams}, E.~R., {Ciardi}, D.~R., {Dupree}, A.~K., {et~al.} 2013, \aj, 146, 71

\bibitem[{{Aitken}(1904)}]{Aitken1904}
{Aitken}, R.~G. 1904, Lick Observatory Bulletin, 3, 6

\bibitem[{{Albrecht} {et~al.}(2012){Albrecht}, {Winn}, {Johnson}, {Howard},
  {Marcy}, {Butler}, {Arriagada}, {Crane}, {Shectman}, {Thompson}, {Hirano},
  {Bakos}, \& {Hartman}}]{Albrecht12}
{Albrecht}, S., {Winn}, J.~N., {Johnson}, J.~A., {et~al.} 2012, \apj, 757, 18

\bibitem[{{Armstrong} {et~al.}(2016){Armstrong}, {de Mooij}, {Barstow},
  {Osborn}, {Blake}, \& {Saniee}}]{Armstrong16}
{Armstrong}, D.~J., {de Mooij}, E., {Barstow}, J., {et~al.} 2016, Nature
  Astronomy, 1, 0004

\bibitem[{{Barnes}(2009)}]{Barnes09}
{Barnes}, J.~W. 2009, \apj, 705, 683

\bibitem[{{Barnes} {et~al.}(2011){Barnes}, {Linscott}, \& {Shporer}}]{Barnes11}
{Barnes}, J.~W., {Linscott}, E., \& {Shporer}, A. 2011, \apjs, 197, 10

\bibitem[{{Benomar} {et~al.}(2014){Benomar}, {Masuda}, {Shibahashi}, \&
  {Suto}}]{Benomar14}
{Benomar}, O., {Masuda}, K., {Shibahashi}, H., \& {Suto}, Y. 2014, \pasj, 66,
  94

\bibitem[{{Borucki} {et~al.}(2011){Borucki}, {Koch}, {Basri}, {Batalha},
  {Brown}, {Bryson}, {Caldwell}, {Christensen-Dalsgaard}, {Cochran}, {DeVore},
  {Dunham}, {Gautier}, {Geary}, {Gilliland}, {Gould}, {Howell}, {Jenkins},
  {Latham}, {Lissauer}, {Marcy}, {Rowe}, {Sasselov}, {Boss}, {Charbonneau},
  {Ciardi}, {Doyle}, {Dupree}, {Ford}, {Fortney}, {Holman}, {Seager},
  {Steffen}, {Tarter}, {Welsh}, {Allen}, {Buchhave}, {Christiansen}, {Clarke},
  {Das}, {D{\'e}sert}, {Endl}, {Fabrycky}, {Fressin}, {Haas}, {Horch},
  {Howard}, {Isaacson}, {Kjeldsen}, {Kolodziejczak}, {Kulesa}, {Li}, {Lucas},
  {Machalek}, {McCarthy}, {MacQueen}, {Meibom}, {Miquel}, {Prsa}, {Quinn},
  {Quintana}, {Ragozzine}, {Sherry}, {Shporer}, {Tenenbaum}, {Torres},
  {Twicken}, {Van Cleve}, {Walkowicz}, {Witteborn}, \& {Still}}]{Borucki11}
{Borucki}, W.~J., {Koch}, D.~G., {Basri}, G., {et~al.} 2011, \apj, 736, 19

\bibitem[{{Brown} {et~al.}(2011){Brown}, {Latham}, {Everett}, \&
  {Esquerdo}}]{Brown11}
{Brown}, T.~M., {Latham}, D.~W., {Everett}, M.~E., \& {Esquerdo}, G.~A. 2011,
  \aj, 142, 112

\bibitem[{{Eastman} {et~al.}(2012){Eastman}, {Gaudi}, \& {Agol}}]{Eastman12}
{Eastman}, J., {Gaudi}, B.~S., \& {Agol}, E. 2012, ascl:1207.001

\bibitem[{{Espinosa Lara} \& {Rieutord}(2011)}]{Espinosa11}
{Espinosa Lara}, F., \& {Rieutord}, M. 2011, \aap, 533, A43

\bibitem[{{Esteves} {et~al.}(2013){Esteves}, {De Mooij}, \&
  {Jayawardhana}}]{Esteves13}
{Esteves}, L.~J., {De Mooij}, E.~J.~W., \& {Jayawardhana}, R. 2013, \apj, 772,
  51

\bibitem[{{Esteves} {et~al.}(2015){Esteves}, {De Mooij}, \&
  {Jayawardhana}}]{Esteves15}
{Esteves}, L.~J., {De Mooij}, E.~J.~W., \& {Jayawardhana}, R. 2015, \apj, 804,
  150

\bibitem[{{Foreman-Mackey} {et~al.}(2013){Foreman-Mackey}, {Hogg}, {Lang}, \&
  {Goodman}}]{Foreman2013}
{Foreman-Mackey}, D., {Hogg}, D.~W., {Lang}, D., \& {Goodman}, J. 2013, \pasp,
  125, 306

\bibitem[{{Holt}(1893)}]{Holt1893}
{Holt}, J.~R. 1893, Astronomy and Astro-Physics (formerly The Sidereal
  Messenger), 12, 646

\bibitem[{{Howarth} \& {Morello}(2017)}]{Howarth17}
{Howarth}, I.~D., \& {Morello}, G. 2017, \mnras, 470, 932

\bibitem[{{Huber} {et~al.}(2014){Huber}, {Silva Aguirre}, {Matthews},
  {Pinsonneault}, {Gaidos}, {Garc{\'{\i}}a}, {Hekker}, {Mathur}, {Mosser},
  {Torres}, {Bastien}, {Basu}, {Bedding}, {Chaplin}, {Demory}, {Fleming},
  {Guo}, {Mann}, {Rowe}, {Serenelli}, {Smith}, \& {Stello}}]{Huber14}
{Huber}, D., {Silva Aguirre}, V., {Matthews}, J.~M., {et~al.} 2014, \apjs, 211,
  2

\bibitem[{{Johnson} {et~al.}(2014){Johnson}, {Cochran}, {Albrecht},
  {Dodson-Robinson}, {Winn}, \& {Gullikson}}]{Johnson14}
{Johnson}, M.~C., {Cochran}, W.~D., {Albrecht}, S., {et~al.} 2014, \apj, 790,
  30

\bibitem[{{Kipping}(2013)}]{Kipping2013}
{Kipping}, D.~M. 2013, \mnras, 435, 2152

\bibitem[{{Lucy}(1967)}]{Lucy1967}
{Lucy}, L.~B. 1967, \zap, 65, 89

\bibitem[{{Mandel} \& {Agol}(2002)}]{MandelAgol02}
{Mandel}, K., \& {Agol}, E. 2002, \apjl, 580, L171

\bibitem[{{Masuda}(2015)}]{Masuda15}
{Masuda}, K. 2015, \apj, 805, 28

\bibitem[{{McLaughlin}(1924)}]{McLaughlin24}
{McLaughlin}, D.~B. 1924, \apj, 60, doi:10.1086/142826

\bibitem[{{Radigan} {et~al.}(2012){Radigan}, {Jayawardhana}, {Lafreni{\`e}re},
  {Artigau}, {Marley}, \& {Saumon}}]{Radigan12}
{Radigan}, J., {Jayawardhana}, R., {Lafreni{\`e}re}, D., {et~al.} 2012, \apj,
  750, 105

\bibitem[{{Rajan} {et~al.}(2015){Rajan}, {Patience}, {Wilson}, {Bulger}, {De
  Rosa}, {Ward-Duong}, {Morley}, {Pont}, \& {Windhorst}}]{Rajan15}
{Rajan}, A., {Patience}, J., {Wilson}, P.~A., {et~al.} 2015, \mnras, 448, 3775

\bibitem[{{Rossiter}(1924)}]{Rossiter24}
{Rossiter}, R.~A. 1924, \apj, 60, doi:10.1086/142825

\bibitem[{{Schlesinger}(1910)}]{Schlesinger1910}
{Schlesinger}, F. 1910, Publications of the Allegheny Observatory of the
  University of Pittsburgh, 1, 123

\bibitem[{{Shporer} {et~al.}(2014){Shporer}, {O'Rourke}, {Knutson},
  {Szab{\'o}}, {Zhao}, {Burrows}, {Fortney}, {Agol}, {Cowan}, {Desert},
  {Howard}, {Isaacson}, {Lewis}, {Showman}, \& {Todorov}}]{Shporer14}
{Shporer}, A., {O'Rourke}, J.~G., {Knutson}, H.~A., {et~al.} 2014, \apj, 788,
  92

\bibitem[{{Szab{\'o}} {et~al.}(2012){Szab{\'o}}, {P{\'a}l}, {Derekas}, {Simon},
  {Szalai}, \& {Kiss}}]{Szabo12}
{Szab{\'o}}, G.~M., {P{\'a}l}, A., {Derekas}, A., {et~al.} 2012, \mnras, 421,
  L122

\bibitem[{{Szab{\'o}} {et~al.}(2014){Szab{\'o}}, {Simon}, \& {Kiss}}]{Szabo14}
{Szab{\'o}}, G.~M., {Simon}, A., \& {Kiss}, L.~L. 2014, \mnras, 437, 1045

\bibitem[{{Szab{\'o}} {et~al.}(2011){Szab{\'o}}, {Szab{\'o}}, {Benk{\H o}},
  {Lehmann}, {Mez{\H o}}, {Simon}, {K{\H o}v{\'a}ri}, {Hodos{\'a}n},
  {Reg{\'a}ly}, \& {Kiss}}]{Szabo11}
{Szab{\'o}}, G.~M., {Szab{\'o}}, R., {Benk{\H o}}, J.~M., {et~al.} 2011, \apjl,
  736, L4

\bibitem[{{Van Eylen} {et~al.}(2013){Van Eylen}, {Lindholm Nielsen}, {Hinrup},
  {Tingley}, \& {Kjeldsen}}]{VanEylen13}
{Van Eylen}, V., {Lindholm Nielsen}, M., {Hinrup}, B., {Tingley}, B., \&
  {Kjeldsen}, H. 2013, \apjl, 774, L19

\bibitem[{{Von Zeipel}(1924)}]{VonZeipel24}
{Von Zeipel}, H. 1924, \mnras, 84, 665

\bibitem[{{Wilson} {et~al.}(2014){Wilson}, {Rajan}, \& {Patience}}]{Wilson14}
{Wilson}, P.~A., {Rajan}, A., \& {Patience}, J. 2014, \aap, 566, A111

\bibitem[{{Winn} {et~al.}(2007){Winn}, {Holman}, {Henry}, {Roussanova}, {Enya},
  {Yoshii}, {Shporer}, {Mazeh}, {Johnson}, {Narita}, \& {Suto}}]{Winn07}
{Winn}, J.~N., {Holman}, M.~J., {Henry}, G.~W., {et~al.} 2007, \aj, 133, 1828

\end{thebibliography}

\end{document}